\RequirePackage{fix-cm}
\documentclass[smallextended]{svjour3}       
\smartqed  
\usepackage{graphicx}
\usepackage{bm} 
\usepackage{amssymb} 
\usepackage{amsmath} 
\begin{document}

\title{Quantum turbulence in coflow of superfluid ${}^4${He} 
}


\author{S. Ikawa         \and
        M. Tsubota 
}


\institute{S. Ikawa \and M.Tsubota \at
              Department of Physics, Osaka City University, Sugimoto 3-3-138, Sumiyoshi-ku 
Osaka 558-8585, Japan \\
              \email{s.ikawa@sci.osaka-cu.ac.jp}           
           \and
           M.Tsubota \at
              The OCU Advanced Research Institute for Natural Science and Technology (OCARINA), Osaka, Japan
}

\date{Received: date / Accepted: date}

\maketitle

\begin{abstract}
We study numerically nonuniform quantum turbulence of coflow in a square channel by the vortex filament model. Coflow means that superfluid velocity $\bm{v}_s$ and normal fluid velocity $\bm{v}_n$ flow in the same direction. Quantum turbulence for thermal counterflow has been long studied theoretically and experimentally. In recent years, experiments of coflow are performed to observe different features from thermal counterflow. By supposing that $\bm{v}_s$ is uniform and $\bm{v}_n$ takes the Hagen-Poiseiulle profile, our simulation finds that quantized vortices are distributed inhomogeneously. Vortices like to accumulate on the surface of a cylinder with $\bm{v}_s \simeq \bm{v}_n$. Consequently, the vortex configuration becomes degenerate from three-dimensional to two-dimensional. 
\end{abstract}

\section{Introduction}
\label{intro}
Quantum turbulence is one of the most important issues in low temperature physics and has been studied theoretically and experimentally  for thermal counterflow where superfluid and normal fluid flow oppositely. In recent years, experiments of coflow where superfluid and normal fluid flow in the same direction are performed to observe different features from thermal counterflow \cite{Ref1}. For example, the vortex line density is proportional to the 3/2 power of the velocity and independent of temperature. The motivation of the present paper is to find numerically some behavior of vortices characteristic of coflow. 

According to the two-fluid model, superfluid ${}^4${He} at a finite temperature consists of an intimate mixture of a viscous normal fluid component (density $\rho_n$ and velocity $\bm{v}_n$) and an inviscid superfluid components (density $\rho_s$ and velocity $\bm{v}_s$) \cite{Ref2}. 
In superfluid ${}^4${He} the circulation of a superfluid vortex, called a quantized vortex, is quantized by the quantum circulation $\kappa=h/m_4$, where $h$ is Planck's constant and $m_4$ is the mass of a ${}^4${He} atom. Quantum turbulence generally occurs by tangling of quantized vortices \cite{Ref3}.  

We perform numerical simulation for coflow in a square channel; superfluid velocity is prescribed to be uniform flow and normal fluid velocity to be Hagen-Poiseuille flow. Vortices are distributed inhomogeneously and like to accumulate through the mutual friction on the surface of a cylinder where superfluid velocity equals normal fluid velocity. As a result, the vortex configuration becomes degenerate from three-dimensional to two-dimensional. How strongly the vortices accumulate depends on temperature and the averaged velocity.

The contents of this paper are as follows. In Sec.~\ref{formulation}, we shall clarify the formulation of the model and introduce the equation of motion. In Sec.~\ref{sec:1}, we show the results for coflow. Section~\ref{sec:3} is devoted to the conclusion and the future work. 

\section{Formulation}
\label{formulation}
We perform numerical simulation by using the vortex filament model with the full Biot-Savart law \cite{Ref4,Ref5}.
A point $\bm{s}$ on the vortex filament is represented in a parametric form  $\bm{s}=\bm{s}(\xi,t)$, where $t$ is time and $\xi$ is the one-dimensional coordinate along the filament. The equation of motion of $\bm{s}$ is given by
\begin{equation}
\dot{\bm{s}}=\bm{v}_s+\alpha\bm{s}'\times(\bm{v}_n-\bm{v}_s)-\alpha'\bm{s}'\times[\bm{s}'\times(\bm{v}_n-\bm{v}_s)],
\end{equation}
where $\alpha$ and $\alpha'$ are the temperature-dependent mutual friction coefficients, and the prime denotes derivatives of $\bm{s}$ with respect to $\xi$. 

The first term of the right-hand side of Eq.(1) is superfluid velocity, given by 
\begin{equation}
\bm{v}_s=\bm{v}_{s,\omega}+\bm{v}_{s,b}+\bm{v}_{s,a}.
\end{equation}
Here $\bm{v}_{s,\omega}$ is the velocity field caused by vortex filaments, $\bm{v}_{s,b}$ the boundary induced field, $\bm{v}_{s,a}$ the applied uniform velocity field. The velocity field $\bm{v}_{s,\omega}$ is represented by the Biot-Savart law
\begin{equation}
\bm{v}_{s,\omega}(\bm{r})=\frac{\kappa}{4\pi}\int_{\cal L}\frac{(\bm{s}_1-\bm{r})\times d\bm{s}_1}{|\bm{s}_1-\bm{r}|^3},
\end{equation}
where $\bm{s}_1$ refers to a point on the vortex filament and the integration is performed along the vortex filaments. The velocity field $\bm{v}_{s,\omega}$ is obtained by a simple procedure; it is just the field produced by an image vortex that is constructed by reflecting the filament into the surface and reversing its direction.

The second and third terms of the right-hand side of Eq.(1) are caused by mutual friction. 
The second term makes a vortex balloon out or collapse inward. 
As discussed in \cite{Ref6},  when the relative velocity $\bm{v}_{ns}=\bm{v}_n-\bm{v}_{s,a}$ flows against $\bm{v}_{s,\omega}$, the mutual friction always shrinks the vortex line locally. On the other hand, $\bm{v}_{ns}$ flowing along $\bm{v}_{s,\omega}$ yields a critical radius of curvature $R_c$. When the local radius $R$ at a point on the vortex is smaller than $R_c$, the vortex shrinks locally, while the vortex balloons out when $R>R_{c}$.

We prescribe the Hagen-Poiseuille profile $u_p$ for $\bm{v}_n$, though the normal fluid flow in the experiment \cite{Ref1} may be actually turbulent. When the normal fluid flows along the ${\it x}$ direction, the ${\it x}$ component of $\bm{v}_n$ is given by 
\begin{equation}
u_p(y,z)=u_0\sum_{m=1,3,5,...}^{\infty}(-1)^{(m-1)/2}\left[1-\frac{\cosh(m\pi z/2a)}{\cosh(m\pi b/2a)}\right]\frac{\cos(m\pi y/2a)}{m^3},
\end{equation}
where $u_0$ is a normalization factor and $a$ and $b$ are halves of the channel width along the $y$ and $z$ axes, respectively \cite{Ref7}.

To characterize the development of vortices, we introduce the vortex line density (VLD) as 
\begin{equation}
L = \frac{1}{\Omega}\int_{\cal L}d\xi,
\end{equation}
where the integral is performed along all vortices in the sample volume $\Omega$. 

In this study, our calculation is performed under the following conditions. A vortex filament is represented by a string of discreet points. The numerical space resolution, namely the minimum distance between neighboring points, is $\Delta\xi=8.0\times10^{-4}$ cm, the time resolution is $\Delta=1.0\times10^{-4}$ s, and the computing box is $0.1\times0.1\times0.1$ ${\rm cm}^3$. 
We consider the case of $\bar{\bm{v}_n}=\bar{\bm{v}_s}$, where $\bar{\bm{v}_n}$ and $\bar{\bm{v}_s}$ are the spatially averaged normal fluid velocity and the spacial average of the applied superfluid velocity $\bm{v}_{s,a}$, respectively. The periodic boundary conditions are used along the flow direction $x$, whereas the solid boundary conditions are applied to channel walls. The effects of reconnection is artificially performed, whenever two vortices approach more closely than $\Delta\xi$. The initial state consists of eight randomly oriented vortex rings of radius 0.023 cm (Fig.\ref{fig:3}[a]). 

\section{Results and Discussion}
 \label{sec:1}

\subsection{Distribution of localized vortices and development of vortex }
 \label{sec:a}
 
\begin{figure}[t]
\begin{center}
  \includegraphics[width=0.97\textwidth]{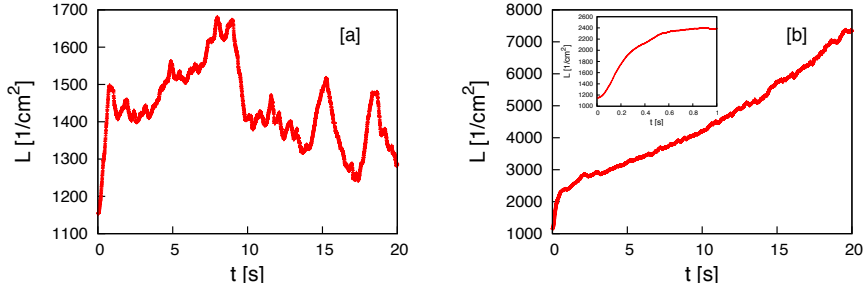}
\caption{Temporal development of the VLD shows two different kinds of behavior depending on the vortex states. In [a] with $T$=1.35K and $\bar{\bm{v}_n}(=\bar{\bm{v}_s})=1.0$ cm/s the VLD oscillates irregularly, when the vortices are diffusive as shown in Fig.\ref{fig:2}[a]. In [b] with $T$=1.95 K and $\bar{\bm{v}_n}=1.0$ cm/s the VLD just increases, when the vortices are localized as shown in Figs.\ref{fig:2}[b] and [c]. (color online)}
\label{fig:1}       
\end{center}
\end{figure}

\begin{figure}[t]
\begin{center}
  \includegraphics[width=1.0\textwidth]{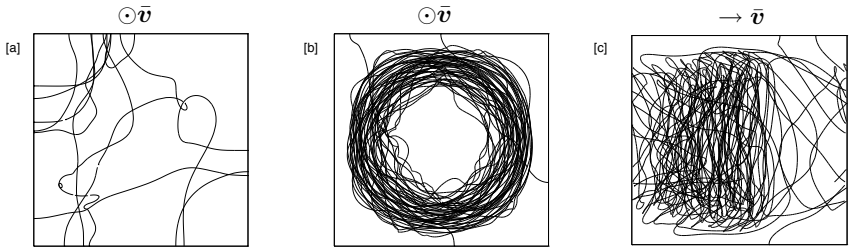}
\caption{Snapshots of vortices viewed along the flow direction; [a] is a snapshot at $t=15$ s of the dynamics of Fig.\ref{fig:1}[a], [b] is a snapshot at $t=20$ s of the dynamics of Fig.\ref{fig:1}[b]. The distribution of vortex has two states, which are diffusive [a] and localized [b]. [c] is the snapshot of [b] viewed from the side.}
\label{fig:2}      
\end{center}
\end{figure}

We perform numerical simulation for coflow in a square channel and find two different states of vortices, namely the diffusive state and the localized state. The features of two states are shown in the figures; Fig.\ref{fig:1} shows the time development of the VLD and Fig.\ref{fig:2} shows the snapshots of the vortices. In the diffusive state, the VLD oscillates irregularly and the vortices are diffusive (Fig.\ref{fig:1}[a] and Fig.\ref{fig:2}[a]). In the localized state, the VLD just increases (Fig.\ref{fig:1}[b]) and the vortices localize in a region shown in Fig.\ref{fig:2}[b],[c].

\begin{figure}[h]
  \begin{minipage}[b]{0.74\textwidth}
  \includegraphics[width=	1.0\textwidth]{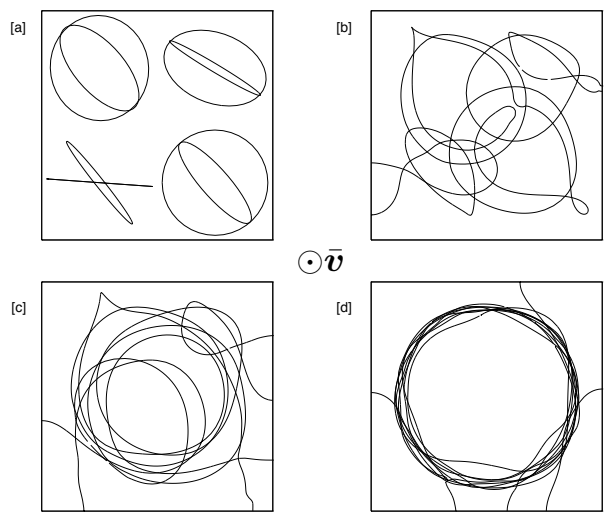}
       \end{minipage}
\begin{minipage}[b]{0.25\textwidth}
\caption{Simulations of the time development of vortex tangle in coflow viewed along the flow direction($T$ = 1.95 K, $\bar{\bm{v}_n}$ = 1.0 cm/s); [a] $t = 0$ s, [b] $t = 0.1$ s, [c] $t = 0.2 $ s, [d] $t = 0.75$ s}
\label{fig:3}
\end{minipage}
\end{figure}

In the localized state, the development of the vortices consists of two stages. As shown in Fig.\ref{fig:1}[b], the first stage is 0 s $\leq t <$ 0.5 s and the second is 0.5 s $\leq t$. In the first stage (Fig.\ref{fig:3}[a],[b],[c]), the vortices repeats lots of reconnections, while they are attracted to a localized region by the mutual friction. Some small vortex loops are made by reconnections and balloon out to a localized region. Then, the VLD increases rapidly. In the second (Fig.\ref{fig:3}[d]), the vortices protruding from the localized region towards the walls continue to extend and wrap the localized region. Since the reconnections seldom occur in this stage, the VLD increases much  slower than that of the first stage.

\begin{figure}[h]
\begin{center}
\includegraphics[width=0.6\textwidth]{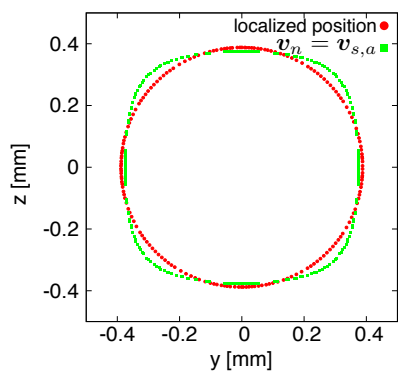}
\caption{The region where mutual friction does not work (color online)}
\label{fig:10}
\end{center}       
\end{figure}

The vortex distribution of coflow is different from that of nonuniform thermal counterflow \cite{Ref8}. This is because that the coflow has the region where the mutual friction dose not work. To see where the vortices accumulate, we consider the localized induction approximation (LIA). Then, the second term of the right-hand side of Eq.(1) is $\alpha \bm{s}'\times(\bm{v}_n-\bm{v}_{s,a}-\beta\bm{s}'\times\bm{s}'')$,
where $\beta$ is the quantity proportional to the quantum circulation \cite{Ref6}. The mutual friction does not work in the region where $\bm{v}_n-\bm{v}_{s,a}-\beta\bm{s}'\times\bm{s}''$ vanishes. For the sake of simplicity, if the term $\beta\bm{s}'\times\bm{s}''$ is negligible, the region is like a square pipe as shown by the square dots in Fig.\ref{fig:10}. However, this region is modified to the round dots by the term $\beta\bm{s}'\times\bm{s}''$; the position around a corner is shifted inward because the radius of curvature is small and the position around a side is shifted  outward because that is large.

\subsection{Parameter dependence of localization}
\label{sec:b}
\begin{figure}[h]
\begin{center}
  \includegraphics[width=0.97\textwidth]{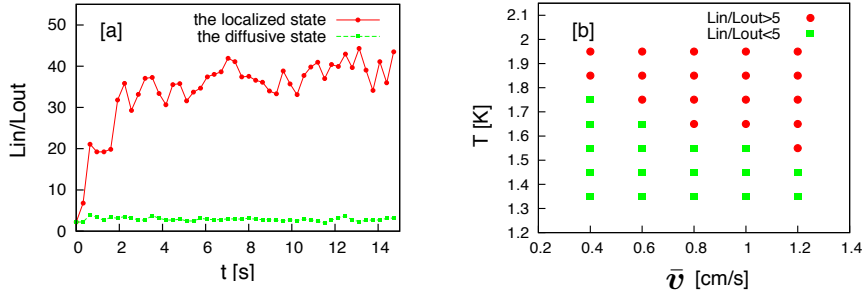}
\caption{[a]: the time development of dimensionless quantity $L_{\rm in}/L_{\rm out}$, [b]: its dependence on temperature and the velocity. In [a], the round dots are for $T=1.95$ K and $\bar{\bm{v}_n}$ = 0.6 cm/s, the square dots are for $T=1.35$ K and $\bar{\bm{v}_n}$ = 1.0 cm/s (color online)}
\label{fig:4}       
\end{center}
\end{figure}

Whether the vortices are localized or diffusive depends on the parameters. We introduce a dimensionless variable $L_{{\rm in}}/L_{{\rm out}}$ to characterize the behavior of the vortices; 
\begin{equation}
\frac{L_{\rm in}}{L_{\rm out}} = \frac{\int_{{\cal L}_{\rm in}}d\xi}{\int_{{\cal L}_{\rm out}}d\xi}.
\end{equation}
Here $L_{\rm in}$ is obtained by the integration along all vortices ${\cal L}_{\rm in}$ in the cylindrical region of the radius $0.045$ cm whose central axis is along that of the pipe and, $L_{\rm out}$ is obtained by the integration along all other vortices ${\cal L}_{\rm out}$. The time development of $L_{{\rm in}}/L_{{\rm out}}$ is shown in Fig.\ref{fig:4}[a]. We define that the vortices are localized when  $L_{{\rm in}}/L_{{\rm out}}$ exceeds 5 after the time passes sufficiently, and that the vortices are diffusive when $L_{{\rm in}}/L_{{\rm out}}$ is less than 5. The value of  $L_{{\rm in}}/L_{{\rm out}}$ depends strongly on temperature and the averaged velocity $\bar{\bm{v}}(=\bar{\bm{v}_n}=\bar{\bm{v}_s})$, and the resulting phase diagram is shown in Fig.\ref{fig:4}[b]. The vortices like to be localized for higher temperature and faster velocity, and diffusive for lower temperature and slower velocity. 

Mutual friction depends on temperature and velocity, because the coefficient $\alpha$ is dependent on temperature and increasing the velocity makes the relative velocity faster everywhere. 
If temperature is lower and velocity is slower, the velocity caused by mutual friction is dominated by the self-induced velocity $\beta\bm{s}'\times\bm{s}''$ and the vortices almost move freely. Consequently, whether the vortices are localized or diffusive depends on temperature and the velocity.

\section{Conclusions}
\label{sec:3}
We found that the vortex development of coflow has two states; one is diffusive and the other is localized. At the localized state, the vortices accumulate on the surface of cylinder where the mutual friction does not work. Whether the vortices are localized or diffusive depends on temperature and the averaged velocity.  

The future work is to perform a numerical simulation of coflow under turbulent normal fluid flow. In the experiment \cite{Ref1}, the normal fluid flow seems turbulent. We will reproduce this situation and confirm this result by calculating the motion of the vortex filament with the turbulent normal fluid flow.

\begin{acknowledgements}
M. T. was supported by JSPS
KAKENHI Grant No. 26400366 and MEXT KAKENHI
"Fluctuation \& Structure" Grant No. 26103526.
\end{acknowledgements}


\begin{thebibliography}{}
%
%
\bibitem{Ref1}
E.Varga, S.Babuin, and L.Skrbek, Physics of Fluids 27, 065101 (2015)
%
\bibitem{Ref2}
R. J. Donnelly, {\it Quantized Vortices in Helium I\hspace{-.1em}I} (Cambridge University Press, Cambridge, 1995), p. 42-45
%
\bibitem{Ref3}
R. P. Feynman, {\it Progress in Low Temperature Physics} (North-Holland, Amsterdam, 1955), Vol. I.
%
\bibitem{Ref4}
K. W. Schwarz, Phys. Rev. B38, 2398 (1988)
%
\bibitem{Ref5}
H. Adachi, S. Fujiyama, and M. Tsubota, Phys. Rev. B 81, 104511 (2010)
%
\bibitem{Ref6}
K. W. Schwarz, Phys. Rev. B31, 5782 (1985)
%
\bibitem{Ref7}
{\it The Handbook of Fluid Dynamics}, (CRC, Boca Raton, 1998)
%
\bibitem{Ref8}
S. Yui, and M. Tsubota, Phys. Rev. B 91, 184504 (2015)
%

\end{thebibliography}


\end{document}